\documentclass[english]{article} 
\usepackage[T1]{fontenc} 
\usepackage[latin1]{inputenc} 
\usepackage{graphicx}
\usepackage{babel} 
\makeatother 
\begin{document}
\title{Interaction  between incoherent light beams propagating in excited atomic hydrogen; applications in astrophysics.
\author{Jacques Moret-Bailly 
\footnote{Laboratoire de physique, Universit\'{e} de Bourgogne, BP 47870,
F-21078
Dijon cedex, France. email : Jacques.Moret-Bailly@u-bourgogne.fr%
}}}

\maketitle
\begin{abstract} 
While it is generally assumed that several light beams propagate independently in a refracting medium, the exception of laser beams may be extended to usual time-incoherent light provided that conditions of space-coherence are fulfilled. Very few molecules have convenient properties, the simplest one being atomic hydrogen in 2S and 2P states (called H* here).

The interaction increases the entropy of a set of beams without a permanent excitation of H*, a loss of energy by a beam having a high Planck's temperature producing a decrease of its frequency, and the thermal radiation getting energy.

Atomic hydrogen in its ground state is pumped to H* by Lyman $\alpha$ absorptions, producing a redshift of the light. The combination of the Lyman absorptions and the redshifts they produce, induce oscillations which generate a spectrum in which the lines deduce from each other by relative frequency shifts which are products of an integer by a constant $z_{b}=0.062$.

\medskip
These purely physical results may be applied in astrophysics, searching where H* may appear. In particular, the computed spectra of the accreting neutron stars, remarkably identical to the spectra of the quasars, may explain that these stars seem never observed. The too high frequencies of the radio signals from the Pioneer probes may result from a transfer of energy from the solar light allowed by a cooling of the solar wind able to produce H*. A similar transfer to the CMB may explain its anisotropy bound to the ecliptic.

\end{abstract}


\section{Introduction}
This is a paper of spectroscopy, not of astrophysics. However, it uses astrophysical objects, observations, or theories as starting points for applications; but the objects are modeled very simply and only very reliable theories are used.

\medskip
Although the first extensively studied light-matter interaction, the refraction is a parametric (i. e. coherent ) effect, attention is generally more paid to interactions of light with individual molecules. The coherent  interaction of a beam of light with a set of molecules, that is an interaction such as the molecular phases (phase of an interacting dipole, for instance) are related identically for all molecules with the local phase(s) of the implied wave(s), is much more powerful than an incoherent interaction because the radiated fields rather than the intensities are added; thus, the total amplitude is multiplied by the square root of the number of molecules, generally very large. Therefore, interactions which are weak at the level of a single molecule may become strong. 

Using lasers or microwaves, a lot of parametric interactions are studied : refraction evidently, photon echoes, phase conjugate reflections, parametric  down conversion, frequency shifts in optical fibres..., the laser itself. But, if several waves are involved, these studies remain topics of specialists because the preservation of the space-coherence in despite of the dispersions requires generally sophisticated equipments.  In some parametric interactions, the state of the matter may be permanently changed. Here, we do not consider the interactions such as the Mössbauer effect after which the matter does not recover its initial state.

\medskip
When a system is perturbed by an electromagnetic wave, its initial stationary state mixes with other states, becoming a  non-stationary ''state of polarisation''. Supposing that the system recovers its initial state after a pulse of light, that is that it is perfectly transparent, the light recovers the energy needed for the polarisation, possibly with some diffraction or change of momentum.  It is the refraction, an effect which is strong although non-resonant because it is always coherent. 

In the semiclassical theory of the refraction by a transparent medium, a refracting molecule excited in a permanent state of polarisation, emits a wave late of $\pi/2$; thus, an input beam $A\sin\omega t$ left nearly unchanged while it crosses a sheet of matter of thickness d$x$, produces, in this sheet, a dynamical polarisation which radiates coherently $Ak$d$x\cos \omega t$; the interference of these lights gives $A\sin(\omega t-k{\rm d}x)$ whose identification with $A\sin(\omega t-2\pi n{\rm d}x/\lambda)$ defines the index of refraction $n$. The refraction may be considered as the result of the interference of the input beam with the beam resulting from virtual two photons interactions with the levels mixed with the initial state.

No blur appears in the refraction because the local phases on the wave surfaces of the exciting wave are related identically to the local phases of the emitted waves.  The parametric exchanges of energy needed by the variations of polarisations are not quantified, the matter working in transient, non stationary states; for instance a beam of the lowest imaginable intensity is refracted by a prism, inducing a transient dynamical polarisation of all molecules of this prism, polarisation which requires a transient, non-quantified change of the energy of the individual molecules. The global state of a set of $N$ refracting molecules, supposed initially in the same non-degenerate state is initially degenerate $N$ times; the polarisation breaks this degeneracy, producing the global state of polarisation which emits the wave late of $\pi/2$, state characterisable by the mode of the exciting light.

\medskip
The hypothesis of transparency does not forbid an exchange of energy between several simultaneous polarisation states of the molecules, therefore between global states of polarisation. The states of molecular polarisation have the same parity, so that they can only interact through a Raman type resonance of frequency $\nu_R$ for instance quadrupolar electric. Considering that a refraction results from a virtual two photons interaction, these exchanges of energy result from four photons virtual interactions, or a set of two simultaneous, virtual Raman transitions (to and from the state which produces the resonance), whose radiations interfere with the incident beams. Using this last model, an elementary computation shows that the changes of energy produce relative changes of frequency $\Delta\nu/\nu$ proportional to  $\nu_R$  and which do not depend on the frequency if the dispersion of the optical constants is neglected \cite{Mor98a}. If the coherence is maintained and the Raman resonance strong, the effect is strong.

\medskip
Various conditions allow to obtain coherent interactions, for instance : i) the sum of the wave vectors of the beams is null or compensated by a global recoil of the medium; ii) in a crystal, the wavelengths of rays of different frequencies may be equal using ordinary and extroardinary propagations;  iii) the light is made of ultrashort pulses \dots. The ultrashort pulses are defined by G. L. Lamb \textquotedblleft shorter than all relevant time constants\textquotedblright \cite{Lamb}; remark that this definition does not involve only the pulses of light, but the medium in which they propagate too. 

Lamb's condition is fulfilled in the optical fibres used for very high speed telecommunications, so that not quantified frequency shifts of the light must be compensated. Although the effect works at any intensity, it is easily observed, thus generally studied, using peak powerful femtosecond lasers, so that the index of refraction, therefore the relative frequency shift depend on the intensity; this effect is named  \textquotedblleft impulsive stimulated Raman scattering''(ISRS) \cite{Yan}. It produces transfers of energy (therefore frequency shifts) from the beams whose temperature deduced from Planck's law is high, to colder beams, often in the thermal range.  These experiments show clearly that the usual assumption that the light beams propagating in matter do not interact, as in the vacuum, is wrong.

\medskip
This assumption is founded on the observation of  the propagation of usual light beams whose interaction named \textquotedblleft Coherent Raman effect on incoherent light'' (CREIL  \footnote{The acronyms ISRS and CREIL are historical, but not very convenient because these effects are far from an ordinary Raman effect: the Raman reference corresponds only to the type of virtual transition which allows a resonance. }) requires a long path in unusual gases:

The usual incoherent light may be considered as made of ultrashort pulses whose length is the coherence time; the CREIL appears if the \textquotedblleft relevant time constants\textquotedblright{} are long enough \cite{Mor98b,Mor01}. An order of magnitude of this coherence time in ordinary light is some nanoseconds, corresponding to some hundreds of megahertz. If Lamb's conditions are fulfilled, energy is transferred between beams of ordinary light

In gases, a first relevant time constant is the collisional time which corresponds evidently to a decoherence. The kinetic theory gives formula allowing a computation of the collisional time, although these formula contain a \textquotedblleft collisional section\textquotedblright{} $\sigma$ whose definition depends on the effect which is studied because there is no sharp distance limit for the interaction between two molecules. 

Very roughly, the collisional time constant is long enough at pressures of the order of 100 Pa.

The wave emitted by a polarisation state perturbed by a Raman interaction may be split into the usual wave late of $\pi/2$ which produces the refraction, and a Raman frequency-shifted wave initially in phase with the exciting wave; the frequency shift introduces beats with the exciting wave, which destroy the coherence of this part of the wave unless the light pulse is shorter enough than the Raman period; this period is a relevant time constant of Lamb's conditions.
Je joins
Unhappily, most molecular periods in molecules are much shorter than some nanoseconds, or appear in excited states whose population is low. Atomic hydrogen in its states of principal quantum number $n=2$ and the selection rule $\Delta F=1$ has frequencies close to 100 Mhz, not too large to forbid the coherence and high enough to give a large CREIL effect ; this excited hydrogen will be named here H*. 

\medskip
Many astrophysicists tried to introduce  \textquotedblleft intrinsic redshifts'' in astrophysics, but they considered interactions of a light beam with a single molecule, so that the transfers of energy were quantified, leading to a destruction of the coherence of the beams, and a blur of the spectra. 

\medskip
The CREIL effect in H* introduces an unusual behaviour of the propagation of light in atomic hydrogen, so that the topic of the paper splits into a purely physical study of the propagation of light in hydrogen (section \ref{chaine}), and into answering the question : Where can it be some H* in the Universe  ? (section \ref{applications}).

Section \ref{chaine} is pure physics while some notations are borrowed from astrophysics; it is a big improvement of a former paper \cite{Mor03} showing how the combination of Lyman alpha absorptions by unexcited atomic hydrogen, and the CREIL produced by the atoms excited by this absorption, may lead to instabilities and the generation of a sort of spectrum. This section is split into subsections corresponding to various macroscopic states of the hydrogen and to various exciting beams. In particular, subsection \ref{VUV} shows that the spectrum of a small, heavy, very hot object embedded in hydrogen is extremely complicated. 

Section  \ref{applications} suggests simple explanations of many astronomical observations, only looking for the presence of H* able to  \textquotedblleft catalyse'' transfers of energy which generally redshift the light and heat the thermal radiation.  

\section{Absorption of a continuous, high frequency spectrum by atomic hydrogen.}\label{chaine}
Over a temperature $T=10 000 K$, les molecules of hydrogen are dissociated. The ionisation energy equals $kT$ for a temperature $T=156 000 K$; as the energy needed for a pumping to the  states of principal quantum number $n=2$ (H* states) is the three fourth of the ionisation energy, it equals $kT$ for $T=117 000 K$. Using Boltzman law, these temperatures may be considered as indicating roughly where these particular states of hydrogen are abundant, remarking however that by a thermal excitation, the proportion of hydrogen in the H* states is much limited by the excitation to higher values of $n$, and by the ionisation at low pressures.

In its ground state (principal quantum number $n=1$) atomic hydrogen has the well known spin recoupling resonance at 1420 MHz, too large to provide coherent frequency shifts by CREIL. In the $n=2$ states, the resonances corresponding to the electric quadrupole allowed transitions ($\Delta F = 1$) have the following frequencies: 178 MHz in the 2S$_{1/2}$ state, 59 MHz in 2P$_{1/2}$ state, and 24 MHz in 2P$_{3/2}$ .The resonances in the more excited states may play a secondary rôle, because the CREIL shifts are proportional to their frequencies which are low.

The column density of hydrogen (in all states) needed to obtain a given redshift (for instance a redshift equal to the Ly$_\alpha$ linewidth) depends on the physical parameters of hydrogen, mainly on the temperature :

\subsection{Very hot hydrogen (T>200 000K).}\label{emission}
At very high temperatures, hydrogen is mainly ionised, without a spectrum, or excited to states of large principal quantum number, practically not active in CREIL. Therefore, there is no frequency shift. If the gas contains impurities, they radiate emission lines of ionised states which are as sharp as the temperature and the pressure allow.

As free protons and electrons do not absorb light, the temperature can be maintained by conduction, convection or electronic heating only. In a quiet system, such temperatures are reached only in thin sheets close to the surface of a very hot object.

\subsection{Excited atomic hydrogen (T $\approx$ 100 000 K).}\label{gap}
Hydrogen is atomic, so that it may be heated to this temperature by absorption  of radiating energy. Remark that an excessive heating ionises the gas, reducing the absorption, so that the temperature is relatively stabilised.
An important fraction of the atoms is in the H* states, so that, if the pressure is not too high, the gas redshifts the light. The simultaneous shift and emission or absorption of lines gives to the lines (written in the spectrum) the width of the shift, so that the absorption is generally low and the lines may be mixed: they cannot be observed. 

\subsection{Atomic hydrogen in its ground state (T $\approx$ 20 000 K).}\label{ground}
The thermal excitation of atomic hydrogen being supposed low, a Ly$_\alpha$ pumping is needed to get H* and a redshift.

\subsubsection{Constant, high UV intensity.}\label{HUV}
Supposing that the UV intensity is large and constant, a partial, constant absorption $\Delta I$ of the intensity at the Lyman $\alpha$ frequency generates H* by a total excitation proportional to $W$, producing a redshift $\Delta\nu$ (fig 1). The initial intensity does not matter, but as the instantaneous intensity of the absorption is proportional to the instantaneous intensity of the light, obtaining the required absorption $\Delta I$ requires a long path if the initial intensity is not much larger than $\Delta I$. Along this long path, a notable amount of H* may be produced by a decay of more excited states, or by the thermal excitation; the CREIL produced by more excited states may play a small role too.
\begin{figure} \begin{center} \includegraphics[height=5 cm] {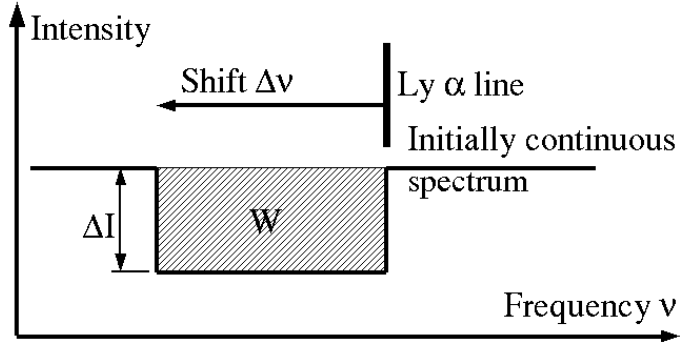}
\end{center}
 \caption{ Permanent absorption by a single line uniformly redshifted of $\Delta\nu$ in atomic hydrogen. For the Lyman alpha line, the pumping proportional to $W= \Delta I \Delta\nu$ provides the  population of H* needed to produce the redshift $\Delta\nu$,  supposing an uniform de-excitation of the excited state.} 
\end{figure}

The constant absorption increases the contrast of a spectrum while the scale of frequencies is changed by the redshift (fig. 2).

\begin{figure} \begin{center} \includegraphics[height=5 cm] {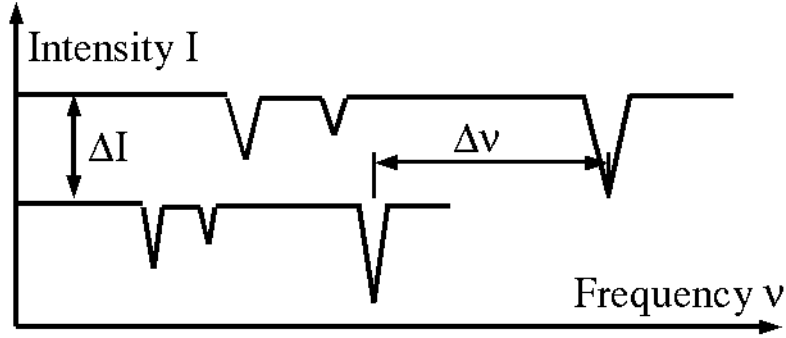}
\end{center}
 \caption{ Absorption of a spectrum and redshift. The final spectrum (low) results from the subtraction of a constant intensity which increases the contrast of the spectrum, and, assuming a constant $\Delta\nu/\nu$, a change of the scale of frequencies.}
 \end{figure}

A stabilisation of the temperature stronger than in \ref{gap} occurs because the redshift which increases the energy available for the Ly$_\alpha$ absorption is reduced up by high thermal excitations and a start of ionisation, down by a dimerisation of the atoms.

\subsubsection{Low UV intensity.}\label{LUV}
If the UV intensity is lower than $\Delta I$, a very long path is necessary both to absorb the available intensity, and to get from the the secondary sources of redshift described in \ref{HUV}
a redshift sufficient to leave the region of low UV intensity.
\subsubsection{Variable UV intensity.}\label{VUV}
Suppose that the intensity is generally high, but that a high frequency absorption line was written into the spectrum where it was no redshift. The spectrum is redshifted as indicated in \ref{HUV} until the absorbed line gets the the Lyman $\alpha$ frequency so that the shift slows down strongly as indicated in \ref{LUV}. During the quasi-stop of redshift, absorptions and emissions of the lines of the gas occur at fixed frequencies of the light, these lines are well written into the spectrum (fig 3). The Ly$_\beta$ and Ly$_\gamma$ are the strongest lines, their frequencies in the light-spectrum may be shifted to the Ly$_\alpha$ frequency, producing new absorbed lines, a multiplication of the absorbed lines.
\begin{figure} \begin{center} \includegraphics[height=10 cm] {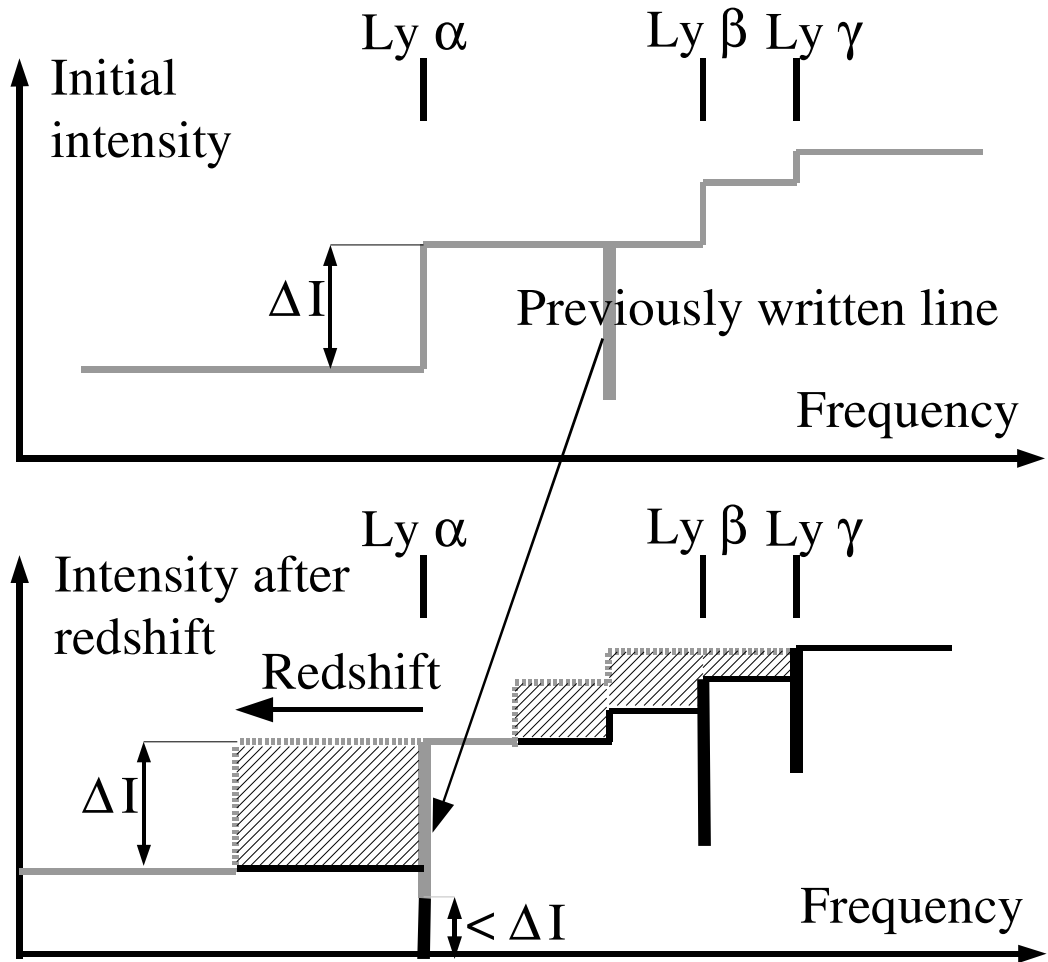} 
\end{center}
\caption{Multiplication of the Lyman spectral lines. The top graph shows a continuous spectrum after an absorption of Lyman lines, and an other line absorbed previously without redshift. During the redshift (low graph) , the hachured regions are absorbed, but the intensity $\Delta I$ cannot be absorbed when the previously written line comes on the Ly$_\alpha$ line, so that the redshift nearly stops and all lines are visibly absorbed. }
\end{figure}

Remark that this process works for a previously written emission line because it produces an acceleration of the redshift, therefore decreases of the absorptions similar to emissions.

The lines observed in the spectrum are more easily characterised by a relative frequency shift $z$ from their initial frequency, than from their final frequencies. Thus, the Ly$_\beta$ (resp.Ly$_\gamma$) line, redshifted to the Ly$_\alpha$ has the redshift:

$$z_{(\beta{\rm {resp.}\gamma,\alpha)}}=\frac{\nu_{(\beta,{\rm resp.}\gamma)}-\nu_{\alpha}}{\nu_{\alpha}}\approx\frac{1-1/(3^{2}{\rm resp.} 4^{2})-(1-1/2^{2})}{1-1/2^{2}}$$
$$z_{(\beta,\alpha)}\approx5/27\approx0.1852\approx3*0.0617;$$
$$z_{(\gamma,\alpha)}=1/4=0.025=4*0.0625.$$
If special conditions allow the $\beta$ line play the same rôle than the $\alpha$:
$$z_{(\gamma,\beta)}\approx7/108\approx0.065.$$
 Notice that the resulting redshifts appear, within a good approximation, as the products of $z_{b}=0.062$ and an integer $q$. The intensities of the Lyman lines are decreasing functions of the final principal quantum number $n$, so that the inscription of a pattern is better for $q=3$ than for $q=4$ and \textit{a fortiori} for $q=1$.

\medskip
 Iterating, the coincidences of the shifted line frequencies with the Lyman $\beta$ or $\alpha$ frequencies build a \textquotedblleft tree\textquotedblright , final values of $q$ being sums of the basic values 4, 3 and 1. Each step being characterised by the value of q, a generation of successive lines is characterised by successive values of $q : q_{1},q_{2}...$ As the final redshift is 
$$q_{F}*z_{b}=(q_{1}+q_{2}+...)*z_{b},$$ the addition $q_{F}=q_{1}+q_{2}+...$ is both a symbolic representation of the successive elementary processes, and the result of these processes. The metaphor \textquotedblleft tree\textquotedblright, is imprecise because \textquotedblleft branches\textquotedblright  {}  of the tree may be \textquotedblleft stacked\textquotedblright {} by coincidences of frequencies. A remarkable coincidence happens for $q=10$, this number is obtained by the effective coincidences deduced from an overlapping sequence of Lyman lines corresponding to the symbolic additions:
$$10=3+3+4=3+4+3=4+3+3=3+3+3+1=... $$
$q=10$ is so remarkable that $z_{f}=10z_{b}=0.62$ may appear a value of $z$ more fundamental than $z_{b}$.

\medskip
In these computations, the levels for a value of the principal quantum number $n$ greater than 4 are neglected, for the simple reason that the corresponding transitions are too weak.

The gas acquires a structure, depending whether the light which crosses it is in a redshift phase, or in a visible line absorption phase.

\subsubsection{Mean absorption.}\label{a}
For low densities, the collisions are negligible during a pulse of light, so that the CREIL depends only on the column density of H*.  For higher densities, leading to collisional times of the order of the coherence time of the light, for a given column density of H* the CREIL decreases fast as the pressure increases.

In thermally unexcited atomic hydrogen, H* is produced mainly by the Lyman $\alpha$ absorption, and destroyed during collisions, the purely spontaneous de-excitation needing a very long time. Thus, for a given column density, if the density decreases, the Lyman absorption is invariant while the CREIL increases. Consequently,  the ratio of the  absorptions to the frequency shifts in the process of subsection \ref{VUV} decreases. Thus, the mean remaining intensity increases.

If the gas cools enough for a dimerisation, the frequencies higher than the Lyman $\alpha$ are absorbed, so that the side effects which allow to leave an absorption phase tend to disappear: there is a large probability to stop the generation of the Lyman forest during an absorption, so that the redshifts are products of $z_{b}$ by an integer.

 \section{Possible applications in astrophysics.}\label{applications}
The CREIL produces  frequency shifts, heatings of the thermal radiation where there is some H* : A good rule should be \textquotedblleft look for H*''. 

Although its previous observation by several authors remains criticised, it seems very remarkable that the constant  $z_{b}=0.062$ has exactly the value found in \ref{VUV}. The CREIL seems able to explain many other observations:

 \subsection{The accreting neutron stars.}\label{accretors}
The theory of the stars is considered as reliable. However, this theory predicts that {\it accretors}, neutron stars accreting a cloud of dirty hydrogen, should be easily observed \cite{Treves,Popov}: although they are small (20 km diameter), they are extremely hot (T> 1 MK), and the size of the cloud produced by the explosion of the generating old and heavy star allows a long life. Precising the spectrum of such a star, we will find properties of quasar observed spectra, therefore give the corresponding references.

The computation of the gravity shows that, in the hypothesis of a static surrounding atmosphere, the thickness of this atmosphere would be between 1 and 10 millimetres. This computation is not reliable because the atmosphere is very dynamical, the gas falling to the surface of the star probably directly, maybe after turns around the star. As the CREIL is not sensitive to the speed of the molecules that it uses as a catalyst, we make only the hypothesis of an atmosphere whose density and temperature decrease with the distance to the surface, without large spatial fluctuations.

The redshifts produced by H*  decrease with the column density of this gas between the source of and the Earth, that is they decrease as the distance from the source to the surface increases.

\medskip
- The surface of the star works as the anticathode of an X rays tube, although the particles which hit it, mainly protons and electrons, have been accelerated by the gravity which is very large because the remaining mass of the small star is of the order of the mass of the Sun. Thus the surface emits electromagnetic waves up to gamma rays. The emissivity of  the surface is not good, so that the temperature of the emitted light is much lower than the temperature of the surface, and than the temperature of the close surrounding gas.

- The very close surrounding gas is extremely hot, strongly ionised, hotter than the radiation from the surface, so that it emits, with the maximum redshift, sharp spectral lines of the impurities, as indicated in \ref{emission}.

- At a larger distance, the gas cools, excited atomic hydrogen appears, which shifts the emissions, the lines disappear, as indicated in \ref{gap}, Therefore, there is a gap of redshifts between the sharp lines and the following ones. This is  observed in quasar spectra. 

- Then the thermal generation of H* almost disappears, and the periodic lines described in \ref{VUV} appear.

\medskip
At pressures of the order of 100 Pa, it remains collisions which widen the lines and lower the redshift; thus the lines of hydrogen and the impurities are broad, very strong, saturated. Saturation means that the light reaches, for the frequencies close to the centre of the lines, the equilibrium with the temperature of the gas, a constant intensity. These broad lines may be emission lines (having a hat shape), or, at a higher altitude, where the temperature of the gas is lower, absorption lines (having shape of a trough). 
However, these pressures favour the acceleration of free electrons by radiowaves, and their collisions with atoms which are ionised: if the star is radio-loud, H* disappears; for a transition, a single broad line is emitted, then absorbed with a nearly constant redshift.

With a decrease of the pressure, it remains only sharper and sharper Lyman lines of hydrogen making a Lyman forest; the mean intensity of the forest increases until the intensity at the Lyman $\alpha$ frequency falls, for instance by an absorption at short wavelengths resulting of a cooling which dimerises the atoms. It is the
 Gunn-Peterson effect \cite{Gunn}. 

\medskip
We obtain a spectrum very similar to the spectrum of a quasar; some accretors are probably seen, but named quasars.  Remark that the identification of quasars with neutron stars is often done for the micro-quasars. The fast speed of the micro-quasars generated in our galaxy could explain their evolution into quasars whose repartition is more isotropic. \cite{Mirabel} 

Some problems which appear in the standard interpretation of the spectrum of a quasar are solved :

* As the cloud was generated by an old star, it may contain abundant heavy elements;

* Supposing that the relative frequency shifts $\Delta\nu/\nu$ are strictly constant, the fine structure patterns appear slightly distorted \cite{Webb}; the dispersion of the optical constants in the CREIL shows that the hypothesis is not strict, so that it is not necessary to suppose that the fine structure constant is a function of the time;

* There is a gap in the redshifts after the sharp emission lines \cite{Rauch,Francis};

* The broad lines which have the shape of troughs do no exist if there is a strong radio emission \cite{Briggs,SAnderson};

* The observed periodicities \cite{Burbidge,Tifft,Hewitt,Bell,Comeau} are simply produced by the propagation of the light in atomic hydrogen. 

A large part of the redshift is intrinsic, as found by Halton Arp \cite{Arp}. Being not extraordinarily far, the quasars are not huge and powerful objects \cite{Petitjean}.

\subsection{Proximity effects.}\label{prox}
* A statistical over abundance of very red objects (VROs) is observed in close proximity to quasars (Hall et al. \cite{Hall}, Wold et al. \cite{Wold}); in particular, the galaxies which contain quasars are often severely reddened, and redshifted relative to other galaxies having similar morphologies (Boller \cite{Boller}). The quasar produces a CREIL redshift, providing far ultraviolet radiation and maybe hydrogen around the VROs. 

\medskip
 * The bright and much redshifted objects seem surrounded by hot dust \cite{Omont}, and it is difficult to explain the stability of this dust in despite of the pressure of radiation and the abrasion by ions. The blueshift, that is the heating of the thermal radiation by the CREIL, as a counterpart of the redshifts in the optical range, is a simple interpretation of the observations.

\subsection{Kotov effect.}
V. A. Kotov and V. M. Lyuty \cite{Lyuty, Kotov} observed oscillations of the luminosity of stars and quasars with a period of 160,01 mn. While the light is redshifted, this period is not. Using CREIL, it is clear that the light pulses are redshifted, but that their starts are not subject to a frequency shift \cite{Lempel}. On the contrary, supposing a change in the scale of time by an expansion of the universe, this result cannot be explained. Therefore, thinking that the observations are reliable, there is no expansion of the universe.

\subsection{Frequency shifts in the solar system.}\label{blue}
Studying the variations of the frequency shifts on the Solar disk allows to compute the fractions due to the Doppler effect and to the gravitation. It remains a redshift proportional to the path of the light through the photosphere, immediately explained by a generation of H* by Lyman pumping.

\medskip
H* may be generated by a combination of the protons and electrons making the Solar wind where the wind cools enough, at the limits of the solar system :
 
               \medskip
* Radio signals were sent from the Earth to Pioneer 10 and 11, at a well stabilised carrier frequency close to 2.11 GHz, and the Pioneers returned a signal after a multiplication of the carrier frequency by 240/221. The blueshift which remains after a standard elimination of the known frequency shifts (Doppler, gravitation) is interpreted as produced by an \textquotedblleft anomalous acceleration\textquotedblright {}(Anderson et al. \cite{Anderson}). 

The CREIL allows to preserve celestial mechanics: The signal obtained during a time constant of a receiver results from the amplitude of the electric field in the mode of reception of the receiving antenna for this time constant; this amplitude results from a weak amplification of the noise (2.7K radiation) by the emission antenna, so that this field is partly incoherent and may be blue-shifted by a CREIL transfer of energy from the light of the Sun.

\medskip
* The anisotropy of the cosmic microwave background \cite{Schwarz} seems bound to the solar system. It may result from an anisotropy of the solar wind observed at it source, in the corona;
when the CMB reaches the solar system its amplification in an anisotropic density of H* may generate its anisotropy .

\medskip
Experiments should be done to study H* in the solar system, in particular changing the intensities and the coherences of the radio signals.

\section{Conclusion}

The CREIL, introduced in previous papers is a parametric effect which, in excited atomic hydrogen H*, transfers energy from electromagnetic modes whose Planck's temperature is high to colder modes, a cooling producing a redshift. Being coherent, the CREIL does not blur the images, and the relative frequency shifts $\Delta\nu/\nu$ are usually nearly constant.

\medskip
A first quantitative result of the CREIL is the computation of the fundamental period $z_{b}=0.062$ of the observed redshifts. Trying to explain observations by the CREIL and by the standard theory, it appears that the CREIL may explain more effects ( frequency shifts of the Pioneer probes, proximity effects, ... ) and does not require extraordinary hypothesis (dark matter, ...).

\medskip
The harsh struggle of some astrophysicists against any evaluation of the CREIL in astrophysics may be a defence of familiar theories, but it  seems more a rejection of all parametric effects justified by the necessity to use, in theories limited to semi-classical models, Planck's quantification rather than Einstein's.


\begin{thebibliography}{10}
\bibitem{Mor98a}Moret-Bailly, J., 1998, Ann. Phys. Fr., 23, C1-235-C1-236
\bibitem{Lamb}Lamb G. L. Jr., 1971, Rev. Mod. Phys., 43, 99-124
\bibitem{Yan} Yan Y.-X., E. B. Gamble Jr. \& K. A. Nelson, 1985, J. Chem Phys., 83, 5391
\bibitem{Mor98b}Moret-Bailly, J., 1998, Quant. \& Semiclas. Opt., 10, L35-L39
\bibitem{Mor01}Moret-Bailly, J., 2001, J. Quant. Spectr. \& Rad. Transfer, 68, 575-582
\bibitem{Mor03}Moret-Bailly, J., 2003, IEEETPS, 31, 1215-1222
\bibitem{Treves}Treves, A. \& M. Colpi, 1991 Astron. Astrophys., 241, 107-111
\bibitem{Popov}Popov, S. B., A. Treves \& R. Turolla, 2003, Astro-ph/0310416
\bibitem{Gunn}Gunn, J. E. \& B. A. Peterson, 1965, ApJ, 142, 1633
\bibitem{Webb}Webb J. K., V. V. Flambaum, C. W. Churchill, M. J. Drinkwater \& J. Barrow, 1999, Phys. Rev. Lett., 82,. 884-887
\bibitem{Mirabel}Mirabel, I. F., I. Rodrigues \& Q. Z. Liu, arxiv:astro-ph/0408562
\bibitem{Rauch}Rauch, M., W. L. W. Sargent, T. A. Barlow, 1999, ApJ, 515, 500-505
\bibitem{Francis}Francis, P. J. \& J. Bland-Hawthorn, arxiv:astro-ph/0405506
\bibitem{Briggs}Briggs, F. H., D. A. Turnshek \& A. M. Wolfe, 1984, ApJ, 287, 549-554
\bibitem{SAnderson}Anderson, S. A., R. J. Weymann, C. B. Foltz \& F. H. Chaffee Jr., 1987,  AJ, 94, 278-288
\bibitem{Burbidge}Burbidge, G., 1968, ApJ., 154, L41-L45
\bibitem{Tifft}Tifft, W. G., 1976, ApJ., 206, 38-56
\bibitem{Hewitt}Burbidge, G. \& A. Hewitt, 1990, ApJ., 359, L33-L36
\bibitem{Bell}Bell, M. B., 2002, Astro-ph/0208320
\bibitem{Comeau}Bell, M. B. \& S. P. Comeau, 2003, Astro-ph/0305060
\bibitem{Arp}Arp H., 2003, Astro-ph/0312198.
\bibitem{Petitjean}Petitjean, P., R. Riediger \& M. Rauch, 1996, A\&A, 307, 417-423
\bibitem{Hall}Hall P. B., M. Sawicki, P. Martini, R. A. Finn, C. P. Pritchett, P. S. Osmer, D. W. McCarthy, A. S. Evans, H. Lin \& F. D. A.
\bibitem{Wold}Wold M., L. Armus, G. Neugegauer, T. H. Jarrett \& M. D. Lehnert, 2003, Astro-ph/0303090
\bibitem{Boller}Boller Th., R. Keil , G. Hasinger, E. Costantini, R. Fujimoto, N. Anabuki, I. Lehmann, L. Gallo, 2003, Astro-ph/0307326 
\bibitem{Omont}Omont, A., R. G. McMahon, J. Bergeron, P.Cox, S. Guilloteau, E. Kreysa, F. Pajot, E. Pecontal, P. Petitjean, P. M. Solomon \& L. J. Storrie-Lombardi, 1997 Early Universe with VLT. Proc. of the ESO workshop, Garching, Germany, 1-4 April 1996. (Springer-verlag), 357-360
\bibitem{Lyuty} Kotov V.A. \& V.M. Lyuty, 1990. CRAS., 310, Ser. II, 743
\bibitem{Kotov} Kotov V.A. 1997. ApJ., 488, 195. 
\bibitem{Lempel} Lempel B. \& J. Moret-Bailly, 2004, unpublished.
\bibitem{Anderson}Anderson, J. D., P. A. Laing, E. L. Lau, A. S. Liu, M. M. Nieto \& S. G. Turyshev, 2002, ArXiv:gr-qc/0104064
\bibitem{Schwarz}Schwarz, D. J., G. D. Starkman, D Huterer, C. J. Copi, Arxiv:Astro-ph/0403353
\end{thebibliography}
\end{document}